\documentclass[aps,prl,preprint,superscriptaddress]{revtex4-1}

\expandafter\let\csname equation*\endcsname=\relax
\expandafter\let\csname endequation*\endcsname=\relaxD
\usepackage{epsf,graphicx,graphics}
\usepackage{amsmath}
\usepackage{latexsym,amssymb}
\usepackage{slashed}
\usepackage{float}
\usepackage{panmaths}

\begin{document}

\newcommand{\xint}{\int\limits_{-1}^{1}}


\title{
	Stable furry black holes in $\suinf$ anti-de Sitter Einstein-Yang-Mills theory, characterised by an infinitude of global charges}


\author{J. Erik Baxter}
\email[]{e.baxter@shu.ac.uk}
\affiliation{Sheffield Hallam University}


\date{\today}

\begin{abstract}
	We present solutions to classical field equations for purely magnetic $\suinf$ Einstein-Yang-Mills theory in asymptotically Anti-de Sitter space. These solutions are found to be stable under linear, time-dependent perturbations. Recent work has also shown that these solutions may in general be uniquely characterized by a countably infinite set of asymptotically measured, gauge-invariant charges. In light of this discovery, we revisit Bizon's `modified No-Hair conjecture', and suggest a new version that accommodates these solutions.
\end{abstract}

\pacs{4.20.Jb}

\maketitle

The study of black hole hair is long and storied. The tale begins with the first ``No-Hair'' uniqueness theorems of Hawking and Israel \cite{israel_event_1967,israel_event_1968,israel_300_1987}, which concluded that static black holes were extremely simple objects and were characterised by 2 parameters: their Arno-Dewitt-Misner (ADM) mass $M$, and their electric charge $e$. A few decades later, infinite families of black holes and solitons were found which fell outside of the remit of these original theorems, in that they were characterised by one extra parameter, the number of zeroes possessed by the single gauge field function $\omega(r)$ \cite{bartnik_particle-like_1988,bizon_gravitating_1994}. Those were unstable, but 
this led Piotr Bizon to suggest the following `modified' No-Hair conjecture \cite{bizon_gravitating_1994}:
\begin{quotation}\noindent
	``Within a given matter model, a stable black hole is uniquely characterised by a finite number of global charges.''
\end{quotation}
Shortly afterwards, stable solutions were found in \emph{asymptotically Anti-de Sitter} (AdS) $\essu$ Einstein-Yang-Mills (EYM) theory \cite{winstanley_existence_1999}, and since then many different generalisations of the original models have been considered, including those for a general gauge group, for dyonic solutions (which also include a non-trivial electric sector), and solutions for non-spherically symmetric spacetimes (see e.g. \cite{baxter_existence_2018}). Some of that work has included models with \emph{infinite} gauge groups -- so called ``furry'' solutions.

Recent work into furry $\suinf$ EYM black holes \cite{baxter_dyonic_2018} demonstrated the existence of solutions which can be globally characterized by a countably infinite set of charges, for large negative values of the cosmological constant $\Lambda$, and this has led us to consider the stability of these models under linear, time-dependent perturbations, in light of Bizon's modified No-Hair conjecture.

\section{Equilibrium field equations}

We begin by reviewing field equations for static, purely magnetic, $\suinf$ EYM field equations. The detailed construction is given in \cite{mavromatos_infinitely_2000,baxter_dyonic_2018}. We use a signature $(-\,,+,+,+)$ and choose units so that $c=G=g=1$ (for $g$ the $\suinf$ coupling constant). We begin with a metric written in standard Schwarzschild co-ordinates $(t,r,\theta,\phi)$:
\begin{equation}\label{metric}
	ds^2=-\mu S^2dt^2+\mu^{-1}dr^2+r^2\left(d\theta^2+\sin^2\theta d\phi^2\right),
\end{equation}
where $\mu$ and $S$ are functions of $r$ alone, and $\mu$ may be written as
\begin{equation}\label{mu}
	\mu(r)=1-\frac{2m(r)}{r}-\frac{\Lambda r^2}{3}.
\end{equation}
Here, $m(r)$ is the \emph{mass function}, and $\Lambda<0$ is the cosmological constant. We use the $\suinf$ gauge potential
\begin{equation}\label{gaugepot}
	\begin{split}
		\mc{W}=\mc{W}_\mu dx^\mu&= i\left[\omega(r,\vartheta)\sin\varphi d\theta\right.\\ &-\left. \left\{\omega(r,\vartheta)\cos\varphi\sin\theta-\cos\vartheta\cos\theta\right\}d\phi\right].
	\end{split}
\end{equation}
We 
emphasise 
the difference between the spacetime angles $(\theta,\phi)$ and the `internal' angles $(\vartheta,\varphi)$ which parametrise the spherical harmonic basis in which $\omega$ is written. 
%
%
It is simpler to transform to the co-ordinate $\xi=\cos\vartheta$, where we notice that $\xi\in[-1,1]$, so we adopt this from now on. With a convenient choice of units, the well-known EYM field equations may be written:
\begin{equation}\label{genFEs}
	\begin{split}
		G_{\mu\nu}+\Lambda g_{\mu\nu}&=3T_{\mu\nu},\\
		\nabla_\mu F^\mu_{\,\,\,\,\nu}+i\left\{\mc{W}_\mu,F^\mu_{\,\,\,\,\nu}\right\}&=0,
	\end{split}
\end{equation}
where the energy-momentum tensor $T_{\mu\nu}$ is 
\begin{equation}
	T_{\mu\nu}=\int_{-1}^1\left(g^{\rho\sigma}F_{\mu\rho}F_{\nu\sigma}-\frac{1}{4}g_{\mu\nu}F^{\rho\sigma}F_{\rho\sigma}\right)d\xi, 
\end{equation}
the anti-symmetric field strength tensor $F_{\mu\nu}$ is given by 
%
	$F_{\mu\nu}=\partial_\mu\mc{W}_\nu-\partial_\nu\mc{W}_\mu+i\left\{\mc{W}_\mu,\mc{W}_\nu\right\}$, 
%
and 
the Poisson bracket is 
%
	$\{f,g\}=\pd{f}{\xi}\pd{g}{\varphi}-\pd{f}{\varphi}\pd{g}{\xi}$. 
%
Substituting the ans\"{a}tze into the field equations, we obtain the following. Writing $^\prime=\frac{d}{dr}$, the Einstein equations become
\begin{equation}\label{EEs}
	\begin{split}
	m^\prime&=\frac{3}{2}\left(\mu G+\frac{P}{r^2}\right),\qquad\frac{S^\prime}{S}=\frac{2G}{r};\\
%
%
		G&=\frac{1}{2}\int_{-1}^1\left(\frac{\partial\omega}{\partial r}\right)^2d\xi,\\
		P&=\frac{1}{8}\int_{-1}^1\left(\frac{\partial}{\partial\xi}\left(\omega^2+\xi^2\right)\right)^2d\xi.
	\end{split}
\end{equation}
The Yang-Mills (YM) equation is given by
%
\begin{equation}\label{YM}
	\begin{split}
		0&=r^2\mu\frac{\partial^2\omega}{\partial r^2}+r^2\left(\mu^\prime+\frac{2\mu G}{r}\right)\frac{\partial\omega}{\partial r}+\frac{\omega}{2}\frac{\partial^2}{\partial\xi^2}\left(\omega^2+\xi^2\right).
	\end{split}
\end{equation}
%
We point out that $S$ decouples from this system, in that it may be calculated once $m$ and $\omega$ have been. Finally, we note that $\omega$, using the properties of Legendre functions, must take the form
\begin{equation}\label{wform}
	\omega(r,\xi)=\sqrt{1-\xi^2}\slim_{j=0}^\infty\omega_j(r)\xi^j.
\end{equation}

Black hole solutions to the field equations \eqref{EEs}, \eqref{YM} are defined on the semi-infinite strip $(r,\xi)\in[r_h,\infty)\times[-1,1]$. We now review the relevant boundary conditions. At the boundary $r=r_h$, where $r_h$ and $\Lambda$ are considered fixed: we find that the mass $m_h$ is fixed by the requirement of a regular event horizon $\mu_h=0$; we rescale $S$ to fix its asymptotic value to be 1, for the correct geometry as $r\rar\infty$, which fixes $S_h$; the gauge field is entirely constrained by the boundary gauge function $\omega_h(\xi)$; and all higher order derivatives of functions at the boundary are then fixed. We have the additional weak constraint on the gauge field boundary values, given by the requirement for a regular non-extremal event horizon: 
\begin{equation}\label{mph}
		\int_{-1}^1\left(\frac{\partial}{\partial\xi}\left(\omega_h^2(\xi)+\xi^2\right)\right)^2d\xi<\frac{8r_h^2}{3}(1-\Lambda r_h^2).
\end{equation}
Asymptotically speaking, we expand all functions in powers of $r^{-1}$. We know that $\lim_{r\rar\infty} S=1$ for the asymptotic AdS limit, and we define $\lim_{r\rar\infty}m(r)=M$, the ADM mass. The freely specifiable asymptotic gauge data are two functions of $\xi$: the asymptotic boundary function $\omega_\infty(\xi)$, and $\Omega(\xi)=-\lim_{r\rar\infty}r^2\pd{\omega}{r}$. All higher order terms in $r^{-1}$ are determined by these two functions and by $M$. Finally, at the boundaries $\xi=\pm1$, it can be seen from \eqref{wform} that $\omega(\pm1)=0\,\forall r$.

We find several trivial solutions to the equilibrium equations \eqref{EEs}, \eqref{YM}. Setting $\omega(r,\xi)\equiv0$, we obtain the Reissner-N\"{o}rdstrom-Anti-de Sitter (RNAdS) solution with mass $m(r)\equiv M$, $S\equiv1$, and global magnetic charge $\mc{Q}^2_M=1$. Setting also $M=0$ here recovers pure AdS space. Setting $\omega(r,\xi)\equiv\sqrt{1-\xi^2}$, we obtain the Schwarzschild-Anti-de Sitter (SAdS) solution also with mass $m(r)\equiv M$ and $S\equiv1$, but with global magnetic charge $\mc{Q}^2_M=0$. There is finally the embedding $\omega(r,\xi)\equiv\omega_*(r)\sqrt{1-\xi^2}$, which it may be seen satisfies the $\essu$ AdS field equations. 

We can also find non-trivial solutions to these field equations in at least two regimes; of interest here are the solutions we find in the limit $|\Lambda|\rar\infty$ \cite{baxter_dyonic_2018}. 
In addition, we obtained an expression for the magnetic charge function, 
%
	$q_M(\xi)=\frac{\sqrt{3}}{2}\pd{}{\xi}\left(\omega_\infty^2+\xi^2\right)$,
%
and we 
showed that in the regime $|\Lambda|\rar\infty$, the function $q_M(\xi)$ uniquely fixes the entire solution (including the ADM mass $M$); and furthermore, $q_M(\xi)$ will in general require an infinite number of gauge field parameters to fully describe it, given \eqref{wform}. Therefore, we now concentrate on solutions in the regime where $|\Lambda|$ is large.


By rescaling the field variables and expanding the functions as power series in $\Lambda^{-1}$ (for $|\Lambda|$ fixed and sufficiently large) \cite{mavromatos_infinitely_2000}, one may prove the existence of solutions (for fixed $r_h$) which exists to all orders in $1/\Lambda$. Additionally, we proved that these solutions are entirely determined by an arbitrary function of $\xi$ which fits the form \eqref{wform}, and which can be taken to be the asymptotic boundary data function $\omega_\infty(\xi)$ \cite{baxter_dyonic_2018}. We there obtain expansions of the form:
\begin{equation}\label{LLexp}
	\begin{split}
		S(r)&=1+O(\Lambda^{-2}),\\
%
		m(r)&=m_h+\frac{\mc{Q}_M^2}{2}\left(\frac{1}{r_h}-\frac{1}{r}\right)+O(\Lambda^{-1}),\\
%
		\omega(r,\xi)&=\omega_\infty(\xi)+O(\Lambda^{-1}),
	\end{split}
\end{equation}
%
%
%
which are given in more detail in \cite{baxter_dyonic_2018}. It is shown that if $\Lambda$ and $r_h$ are fixed, then these expansions, and therefore also the ADM mass $M$, are determined entirely and uniquely (to all orders) by the single asymptotic boundary function $\omega_\infty(\xi)$. 

\section{The time-dependent system}

Next, we describe the analysis of the full non-static field equations. In this case, all functions are upgraded to also be functions of $t$, and we use the following potential, which has been fixed with a `temporal' gauge:
\begin{equation}\label{gaugepot_time}
	\begin{split}
		\mc{W}&=i\left[\frac{1}{2}\mc{B}(t,r,\xi)dr+\omega(t,r,\xi)\sin(\gamma(t,r,\xi)+\varphi) d\theta\right.\\ 
		&-\left.\left\{\omega(t,r,\xi)\cos(\gamma(t,r,\xi)+\varphi)\sin\theta-\xi\cos\theta\right\}d\phi\right].
	\end{split}
\end{equation}
for three real functions $\mc{B}$, $\omega$ and $\gamma$. We then introduce a time-dependent perturbation to all functions, writing them as the sum of a static equilibrium solution and a small time-dependent perturbation quantity, as follows:
\begin{equation}\label{pert_exp}
	\begin{split}
		m(t,r)&=\tilde{m}(r)+\delta m(t,r),\\
		S(t,r)&=\tilde{S}(r)+\delta S(t,r),\\
		\mc{B}(t,r,\xi)&=\delta\mc{B}(t,r,\xi),\\
		\omega(t,r,\xi)&=\tilde{\omega}(r,\xi)+\delta\omega(t,r,\xi),\\
		\gamma(t,r,\xi)&=\delta\gamma(t,r,\xi),
	\end{split}
\end{equation}
and where it is understood that terms of order $\delta^2$ and higher will be ignored. We substitute \eqref{metric}, \eqref{gaugepot_time}, and \eqref{pert_exp} into the field equations \eqref{genFEs}, and note that this form of the gauge potential causes the system to decouple into two sectors: i) the \emph{sphaleronic} sector, containing the gauge field perturbations $\delta\mc{B}$ and $\delta\gamma$, and ii) the \emph{gravitational} sector, containing the gauge field perturbation $\delta\omega$ and the metric perturbations $\delta m$ and $\delta S$. This echoes the stability analysis for $\sun$ \cite{baxter_abundant_2008}. The full system is extremely complicated and the analysis is involved, so we leave the details of the proof for a future work, and here discuss the main results.


We start with the sphaleronic sector. Here, we define the 2-vector $\bsym{\Psi}$ which essentially contains the functions $\delta\mc{B}$ and $\delta\gamma$ though written in a different form. The analysis is lengthy, but tractable, and we obtain the system 
%
	$-\ddot{\bsym{\Psi}}=\mc{U}\bsym{\Psi}$, 
%
where the operator $\mc{U}$ is a second-order partial differential matrix operator in the variables $r$ and $\xi$. The requirement that this system possesses solutions with no unstable modes produces the following condition for stability, which is valid for a general value of $\Lambda<0$:
\begin{equation}\label{stabcon}
	\pd{^2}{\xi^2}\left(\tilde{\omega}^2+\xi^2\right)\leq 0,\,\,\,\forall r,\xi,
\end{equation}
for $\tilde{\omega}(r,\xi)$ the gauge function of the \emph{equilibrium} solution about which the perturbation is being taken. In \cite{baxter_dyonic_2018}, we show that one may recover the $\sun$ field equations from the $\suinf$ field equations, employing a `method of lines' in addition to rescaling all the variables. It is pleasing to note that if we apply this method to \eqref{stabcon}, we recover the $N-1$ conditions 
%
	$\omega^2_j(r)\geq1+\frac{1}{2}\left(\omega_{j-1}^2(r)+\omega_{j+1}^2(r)\right)$, 
%
which are exactly the conditions for stability for the sphaleronic sector in the $\sun$ case \cite{baxter_stability_2016}.


As for the gravitational sector, we can determine that the metric perturbation $\delta S$ again decouples, and after a lot of algebra, $\delta m$ also decouples due to asymptotic regularity requirements. Therefore we can concentrate on the gauge sector, which reduces to 
%
$-\ddot{\delta\omega}=\partial^2_{r_*}\delta\omega+\mc{M}\delta\omega$, 
%
where $\mc{M}$ is a second-order partial differential scalar operator and $r_*$ is a `tortoise' co-ordinate. This sector is a lot more resistant to direct analysis than the sphaleronic. However, in the limit $|\Lambda|\rar\infty$, we can see from \eqref{LLexp} that 
$\pd{\tilde{\omega}}{r}\rar 0$, making the analysis a lot easier; for instance, $\delta m,\delta S\rar0$. Then we obtain the same stability condition as in the sphaleronic sector \eqref{stabcon} -- again this is identical to the situation in $\sun$ \cite{baxter_stability_2016}. Hence, it just remains to show that there exist at least some equilibrium solutions, for some large value of $|\Lambda|$, which conform to the conditions \eqref{mph} and \eqref{stabcon}.


\section{An example solution} 

Here we show a `toy model' of a non-trivial equilibrium gauge field function $\tilde{\omega}(r,\xi)$ which in the limit $|\Lambda|\rar\infty$ fulfils the stability criterion \eqref{stabcon}. We choose the asymptotic boundary data for $\tilde{\omega}$ to be
%
%
%
\begin{equation}\label{eg}
	\tilde{\omega}_\infty(\xi)=\frac{W}{\sqrt{2}}\sqrt{1-\xi^2}\sqrt{1+\frac{\cos\left(\frac{\pi\xi}{2}\right)}{1-\xi^2}},
\end{equation}
which clearly obeys \eqref{wform}, if we expand the final factor in \eqref{eg} as a MacLaurin series in $\xi$. Doing this obtains an infinite set of constants $\tilde{\omega}_{\infty,j}$ which fix $\tilde{\omega}_\infty(\xi)$ -- this shows that the solution is non-trivial. The constant parameter $W$ controls the magnitude of $\tilde{\omega}_\infty$, in the sense that $W=\tilde{\omega}_\infty(0)=\max_{\xi\in[-1,1]}\omega_\infty(\xi)$. 
We substitute \eqref{eg} into the expansions \eqref{LLexp}, with the aim of finding some which meet the stability criterion to all orders of $\Lambda^{-1}$, for some large, fixed value of $|\Lambda|$.

\begin{figure}
	\includegraphics[scale=0.7]{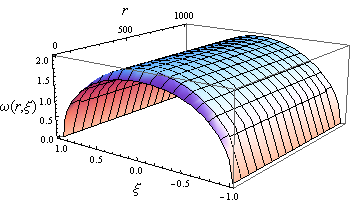}
	\caption{A plot of $\tilde{\omega}(r,\xi)$ for the non-trivial $\suinf$ solution defined by \eqref{LLexp} and \eqref{eg}, with $W=2$, $r_h=1$ and $\Lambda=-20$, including terms up to $O(\Lambda^{-2})$. We call the solution with these parameters `Solution (A)'.\label{BH_eg}}
\end{figure}

\begin{figure}
	\includegraphics[scale=0.6]{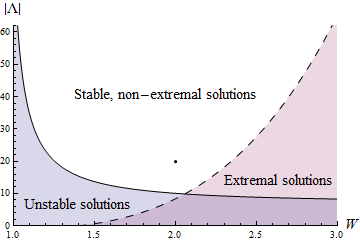}
	\caption{A parameter space plot of $|\Lambda|$ \emph{vs.} $W$ for the solutions defined by \eqref{LLexp} and \eqref{eg}, with $r_h=1$. The area \emph{above} the shaded portions is the region where stable solutions with non-extremal event horizons may be found. The point marks where Solution (A) is located. \label{LvW}}
\end{figure}

Figure \ref{BH_eg} shows a plot of the gauge field $\tilde{\omega}(r,\xi)$ which is defined by \eqref{LLexp} and \eqref{eg} with $W=2$, $r_h=1$ and $\Lambda=-20$, where we include terms up to and including $O(\Lambda^{-2})$. We call this `Solution (A)'. We only show to up to $r=1000$ but the asymptotic behaviour is already apparent by this point.  
To be sure that these solutions were valid in the sense of the convergence of the series \eqref{LLexp}, we could choose a much larger value of $|\Lambda|$, but this would obscure the features of the plots which are qualitatively the same for any value of $|\Lambda|\gtrsim10$. Figure \ref{LvW} shows a parameter space plot of $|\Lambda|$ against $W$. The region above the full line is where $\max_{\forall r,\xi}\left[\pd{^2}{\xi^2}(\tilde{\omega}^2+\xi^2)\right]<0$ \eqref{stabcon}, and the region above the dashed line is where the requirement for a regular event horizon \eqref{mph} is satisfied. 
The plotted point marks Solution (A). We note that we have the absolute lower bound $|\Lambda|=9.89$ (to 3 s.f.) for solutions to meet both the criteria \eqref{mph} and \eqref{stabcon}. Also, as in the $\sun$ case \cite{baxter_soliton_2007}, we see that in the limit $|\Lambda|\rar\infty$, the range of $W$ in which we find stable solutions with regular event horizons grows without limit. 

\begin{figure}
	\includegraphics[scale=0.6]{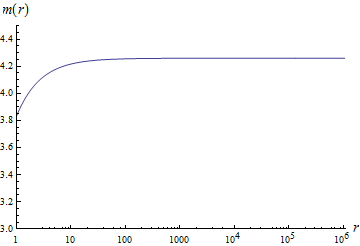}
	\caption{A plot of $m(r)$ for Solution (A), including terms up to $O(\Lambda^{-2})$. One can see that it is monotonically increasing to a limit, as expected, and that the ADM mass is approximately $M=4.26$ (to 3 s.f.).\label{mplot}}
\end{figure}

\begin{figure}
	\includegraphics[scale=0.7]{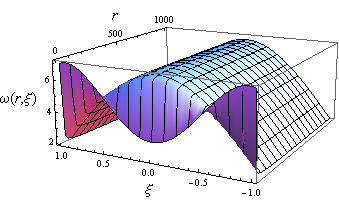}
	\caption{A plot of $-\pd{^2}{\xi^2}\left(\tilde{\omega}^2+\xi^2\right)$ for Solution (A): we note that it is everywhere positive, implying that the solution is stable \eqref{stabcon}.\label{cons}}
\end{figure}

The mass function $m(r)$ of Solution (A) can be seen in Figure \ref{mplot}, and it resembles a typical $\essu$ mass function \cite{winstanley_existence_1999}. 
Figure \ref{cons} is a plot of $-\pd{^2}{\xi^2}\left(\tilde{\omega}^2+\xi^2\right)$ for Solution (A), to double-check the stability condition \eqref{stabcon}; and it can be seen that this is positive everywhere, indicating that the perturbed solution $\omega$ is stable \eqref{stabcon} under the linear time-dependent perturbation \eqref{pert_exp}. 

Above, we have not included terms of $O(\Lambda^{-3})$ or higher, but it can be seen that these terms are extremely small and in any case, if upon inclusion the stability condition \eqref{stabcon} were violated, it is clear that increasing the value of $|\Lambda|$ could again satisfy it by reducing the deviation of $\tilde{\omega}$ from $\tilde{\omega}_\infty$ \eqref{LLexp}. 
%
 The above represents only one possible solution family which we used as a demonstration: many others will exist.


\section{Conclusions}

Here we have presented black hole solutions to $\suinf$ EYM AdS field equations, in the limit where $|\Lambda|$ is large, which are both stable and can be uniquely characterised by no less than a countably infinite number of parameters, which are essentially the MacLaurin coefficients of the asymptotic gauge boundary function $\tilde{\omega}_\infty(\xi)$. Noting the precise wording of Bizon's modified No-Hair conjecture, it is clear that we have discovered at least some solutions which fall \emph{outside} of its scope; the ``matter model'' here being a classical version of a `stringy' black hole -- the field theory limit of an $\suinf$ EYM black hole in the bulk space. 
Therefore, since we are now as far from `\emph{no}-hair' as we can be, the author suggests the following wording for what we venture to call an `Abundant Hair conjecture':
\begin{quotation}\noindent
	``Within a given matter model, a stable black hole can be uniquely characterised by \emph{up to a countably infinite} number of global charges.''
\end{quotation}
%

Since furry EYM solutions are of interest to the AdS/CFT conjecture \cite{ellis_w_infty_2016}, a study of their holographic properties would be interesting, as would results for topological furry black holes, in light of e.g.  \cite{hendi_holographical_2018}. We hope to return to these issues in later work.





\end{document}